\documentclass[%
 reprint,
 superscriptaddress,
 nofootinbib,
 amsmath,amssymb,
 aps,
 pre,
]{revtex4-2}

\usepackage{graphicx}
\usepackage{bm}
\usepackage{color}
\usepackage[colorlinks=true, linkcolor=blue, citecolor=blue, urlcolor=blue]{hyperref}

\makeatletter
\newcounter{subfig}[figure]

\newcommand{\makesubfig}[1]{\refstepcounter{subfig}\label{#1}}
\newcommand{\figsubref}[2]{\hyperref[#2]{\ref*{#1}\ref*{#2}}}
\makeatother

\newcommand\tbbint{{-\mkern -16mu\int}}
\newcommand\dbbint{{-\mkern -19mu\int}}
\newcommand\bbint{\mathop{\mathchoice{\dbbint}{\tbbint}{\tbbint}{\tbbint}\mkern -6mu}\nolimits}

\definecolor{midbluepurple}{RGB}{134,0,102}

\begin{document}

\title{Computing stable configurations of confined smectic liquid crystals with a deep variational framework}

\author{Yuchen Xie}
\affiliation{School of Mathematical Sciences, Peking University, Beijing 100871, China}

\author{Baoming Shi}
\affiliation{Department of Applied Physics and Applied Mathematics, Columbia University, New York, NY 10027, USA}

\author{Yucen Han}
\affiliation{Center for Applied Mathematics, Renmin University of China, Beijing 100872, China}

\author{Lei Zhang}
\email[Contact author: ]{zhangl@math.pku.edu.cn}
\affiliation{School of Mathematical Sciences, Beijing International Center for Mathematical Research, Center for Quantitative Biology, Center for Machine Learning Research, Peking University, Beijing 100871, China}

\begin{abstract}
Smectic liquid crystals are layered liquid-crystalline phases characterized by orientational order and periodic density modulation. Although their structures can be modeled using continuum theories, computing stable configurations remains challenging in complex geometries, particularly when the high-frequency density modulations associated with smectic layering should be resolved. We propose a deep variational framework (DVF) for computing these configurations within the modified Landau--de Gennes model, in which the coupled orientational and positional order parameters are represented on a regular reference domain while physical confinement is incorporated through coordinate mappings. A warmup penalty mitigates the spectral bias of neural networks toward smooth, nonlayered fields, enabling robust recovery of oscillatory smectic states. Comparisons with a neural-network baseline and finite-difference relaxation demonstrate the essential role of this penalty and the numerical stability of the resulting layered states. The DVF reproduces experimentally established smectic-A defect structures and layer morphologies across diverse confinement geometries and further predicts a chevron-like smectic-C state in a tangent-anchored sphere. Together, these results demonstrate the applicability of the DVF to computing stable smectic configurations across experimentally relevant confinement geometries and anchoring conditions.

\end{abstract}

\maketitle

\section{Introduction}
\label{sec:intro}
Liquid crystals are soft condensed matter systems that combine fluidity with partial molecular order \cite{de1993physics,wang2021modelling}. Among their ordered phases, smectic liquid crystals are of particular interest because, in addition to orientational alignment, they develop a one-dimensional density modulation that organizes the material into layers \cite{de1972analogy}. This layered order makes smectics especially sensitive to confinement, surface anchoring, and geometric frustration, since the preferred layer structure is often difficult to accommodate in bounded or curved domains \cite{shojaei2006role,liang2011nematic,monderkamp2021topology}. As a result, confined smectics can exhibit a rich variety of layer morphologies, defect structures, and metastable states. Two representative phases are smectic-A (SmA), in which the average molecular orientation is parallel to the layer normal, and smectic-C (SmC), in which the molecules are collectively tilted with respect to the layer normal. This distinction can lead to qualitatively different confined structures in the two phases \cite{sharma2024smectic}.

To describe confined smectic liquid crystals within a continuum framework, both orientational and positional order must be resolved, with the latter corresponding to the density modulation associated with layering. The Landau--de Gennes (LdG) theory represents orientational order through a tensor field $\bm{Q}$ \cite{de1993physics}, whereas smectic phases require an additional positional order parameter \cite{chen1976landau,pevnyi2014modeling,xia2021structural}. We therefore adopt the modified Landau--de Gennes (mLdG) model \cite{xia2021structural,xia2023variational,xia2024simple}, in which $\bm{Q}$ is coupled to a scalar positional field $u$. This formulation provides a unified variational description of molecular orientation, layered density structure, and their associated defects \cite{xia2021structural,shi2025amodified}. Within this framework, the confined configurations of interest are characterized as local minimizers of the corresponding free-energy functional.

Computing these minima is challenging because three numerical difficulties occur simultaneously. The confinement may be irregular or three-dimensional; the mLdG energy contains second derivatives of $u$ and therefore leads to a fourth-order Euler--Lagrange (EL) equation \cite{xia2023variational}; and the density field becomes strongly oscillatory when many smectic layers fit inside the domain \cite{wittmann2021particle,wittmann2023colloidal}. Established discretization methods can provide accurate solutions, but accommodating complex geometry, higher-order regularity, and fine layer scales within a single implementation can be technically demanding \cite{xia2021structural,xia2023variational,xia2024simple}. Neural variational methods offer a complementary approach by representing the order parameters with trainable functions and minimizing the free energy directly. A standard neural variational method such as the Deep Ritz method (DRM), however, remains susceptible to spectral bias \cite{yu2018deep,rahaman2019spectral,xu2020frequency}. This bias causes neural networks to learn low-frequency components before high-frequency ones and may consequently lead the optimization to a nearly homogeneous, nonlayered branch.

To address these difficulties, we employ a deep variational framework (DVF) to compute stable layered configurations of the coupled mLdG model in complex domains. The DVF is adapted from our previous work on the Landau--Brazovskii model \cite{xie2026geometry}. The neural network serves as a variational representation of $(u,\bm{Q})$, rather than as a surrogate trained on labeled data, and is evaluated on a regular reference domain. A coordinate mapping incorporates the physical confinement \cite{gao2021phygeonet}, separating the treatment of complex boundaries from the representation of the order parameters. To recover the high-frequency density modulation, a warmup penalty promotes the prescribed smectic wavelength during the early optimization and is subsequently removed, so that the final configuration is governed by the original mLdG energy.

Our numerical results show that the DVF recovers stable layered configurations across diverse confinement settings, reproducing established morphologies while revealing additional geometry-dependent structures. We first assess the method in a square by comparison with the DRM, an ablated DVF without the warmup penalty, and a subsequent finite-difference relaxation. We then apply it to SmA confinement in representative two- and three-dimensional (2D and 3D) geometries. The computed SmA minima reproduce corner-localized defects reported in polygonal continuum studies \cite{han2020reduced,han2021solution}, bipolar textures observed in spherical confinement \cite{lopez2009topological,zhou2023defect}, and concentric layers found in particle simulations under homeotropic anchoring \cite{emerson1995monte,puschel2017impact}. In irregular 2D domains, the calculations further relate polygonal corner types to interior layer junctions or corner-localized dislocations and show that tactoid elongation regularizes the interior layers. In 3D, anchoring designed to induce a toroidal focal conic domain (TFCD) produces a localized near-surface order-reconstruction state, while the tangent-anchored sphere supports chevron-like SmC layers \cite{manyuhina2015thick,sharma2024smectic}. Repeated optimizations also recover competing configurations under the same macroscopic conditions, indicating a frustrated mLdG energy landscape. The model and method are introduced in Secs.~\ref{sec:problems}--\ref{sec:method}, followed by numerical results in Sec.~\ref{sec:results} and conclusions in Sec.~\ref{sec:conclusion}.

\section{Mathematical models}
\label{sec:problems}
To describe confined smectic phases, we adopt the mLdG framework \cite{xia2021structural,xia2023variational,xia2024simple}, a phenomenological continuum model formulated in terms of two coupled order parameters: a symmetric, traceless nematic tensor $\bm{Q}\in\mathbb{R}^{d\times d}$ and a scalar positional field $u\in\mathbb{R}$, where $d=2,3$. The tensor $\bm Q$ characterizes local orientational order: its principal eigenvector defines the director $\bm n$, while its eigenvalues measure the strength and possible biaxiality of the order \cite{de1993physics,mottram2014introduction}. The field $u$ measures the deviation of the molecular density from its spatial average, so its periodic modulation represents the alternating high- and low-density regions common to smectic phases. Locally, a modulation of the form $u(\bm{x})\sim\cos(\bm{k}\cdot\bm{x})$ describes layers with normal $\bm{k}/|\bm{k}|$ and spacing $2\pi/|\bm{k}|$. The coupling terms introduced below encode this preferred smectic layering by selecting the wave number and determining the phase-dependent relation between the layer normal and the director $\bm n$.

We seek local minimizers of the total free-energy functional. Over a physical domain $\Omega$, the total energy is given by
\begin{equation}
E(\bm{Q}, u) = \int_{\Omega} \left( f_{\text{LdG}}(\bm{Q}) + f_{\text{bs}}(u) + f_{\text{int}}(\bm{Q}, u) \right)\,\mathrm{d}\bm{x}.
\label{eq:energy}
\end{equation}
The integrand consists of three contributions: the nematic bulk and elastic energy $f_{\text{LdG}}$, the smectic bulk energy $f_{\text{bs}}$, and the coupling term $f_{\text{int}}$ linking orientational and positional ordering.

The first term in Eq.~\eqref{eq:energy} is the nematic free-energy density,
\begin{equation}
f_{\text{LdG}}(\bm{Q}) = \frac{K}{2} |\nabla \bm{Q}|^2 + f_{\text{bn}}(\bm{Q}),
\label{eq:ldg_density}
\end{equation}
where $K>0$ is the elastic constant and $f_{\text{bn}}$ is the local bulk potential governing the transition between isotropic and nematically ordered states. We use the standard LdG potential in 3D and its reduced counterpart in 2D. Their explicit forms and the corresponding equilibrium scalar order $s_+$ are collected in Appendix~\ref{app:nondim}. The connection between orientational order and the smectic density modulation is introduced through $f_{\text{int}}$ below.

The second term in Eq.~\eqref{eq:energy} is the bulk smectic free-energy density associated with the positional order parameter $u$, defined as
\begin{equation}
f_{\text{bs}}(u) = \frac{a}{2}u^2 + \frac{b}{3}u^3 + \frac{c}{4}u^4,
\label{eq:smectic_bulk_density}
\end{equation}
where $a=\alpha_2(T-T_2^\ast)$ is a temperature-dependent coefficient, with $\alpha_2>0$ and $T_2^\ast<T_1^\ast$ associated with the nematic--smectic transition, while $b$ and $c$ are material-dependent constants with $c>0$. In the present work, we set $b=0$, so the bulk potential is invariant under $u\mapsto-u$ \cite{pevnyi2014modeling}. When $a>0$, the homogeneous state $u=0$ is favored, whereas $a<0$ favors a nonzero density-modulation amplitude. Thus, $f_{\text{bs}}$ controls the onset and amplitude of positional order.

The spatial organization of this positional order is governed by the third term $f_{\text{int}}$, whose phase-dependent form selects the layer wavelength and its relation to $\bm Q$. For the SmA phase, we take
\begin{equation}
f_{\text{int}}(\bm{Q}, u) = B_0 \left| D^2u + q^2 \left( \frac{\bm{Q}}{s_+} + \frac{\bm{I}_d}{d} \right) u \right|^2,
\label{eq:sma_coupling}
\end{equation}
whereas for the SmC phase, we use
\begin{equation}
\begin{aligned}
f_{\text{int}}(\bm{Q}, u) &= B_0 \Bigg( (\Delta u + q^2 u)^2 \\
&\; + \Big( \operatorname{tr} \Big( D^2 u \left( \frac{\bm{Q}}{s_+} + \frac{\bm{I}_d}{d} \right) \Big) + q^2 u \cos^2\theta_0 \Big)^2 \Bigg).
\end{aligned}
\label{eq:smc_coupling}
\end{equation}
Here $B_0>0$ weights the energetic penalty for deviations from the preferred geometric relation between the layer normal and the nematic director: alignment in the SmA phase and a prescribed tilt angle $\theta_0$ in the SmC phase. Moreover, $D^2u$ denotes the Hessian matrix of $u$, $q$ is the characteristic wave number of the smectic layers, $\bm{I}_d$ is the $d$-dimensional identity matrix, and $s_+$ is the equilibrium scalar order selected by $f_{\mathrm{bn}}$.

The preferred layered structure follows directly from the phase-dependent coupling terms in Eqs.~\eqref{eq:sma_coupling} and \eqref{eq:smc_coupling}. For a uniaxial state and a local plane wave,
\begin{equation*}
\bm{Q}
=s_+\left(\bm{n}\otimes\bm{n}-\frac{\bm{I}_d}{d}\right),
\qquad
u(\bm{x})=u_0\cos(\bm{k}\cdot\bm{x}),
\end{equation*}
we have
\begin{equation*}
\frac{\bm{Q}}{s_+}+\frac{\bm{I}_d}{d}
=\bm{n}\otimes\bm{n},
\qquad
D^2u=-(\bm{k}\otimes\bm{k})u.
\end{equation*}
The SmA coupling is minimized locally when
\begin{equation}
\bm{k}=\pm q\bm{n},
\label{eq:sma_wavevector}
\end{equation}
so it selects layers normal to the director with spacing $2\pi/q$. For SmC, the two coupling terms instead impose
\begin{equation}
|\bm{k}|=q,
\qquad
|\bm{k}\cdot\bm{n}|=q\cos\theta_0,
\label{eq:smc_wavevector}
\end{equation}
which selects the same wavelength and an angle $\theta_0$ between the director and layer normal. Thus, the layered modulation of $u$ is selected by the mLdG energy rather than imposed externally.

Throughout, boundary anchoring constrains $\bm Q$, whereas $u$ is varied freely and satisfies the natural boundary conditions of the mLdG energy. The conditions used in the computations are summarized in Appendix~\ref{appd:auxloss}.

Hereafter we work with the dimensionless free-energy functional and omit bars for notational simplicity. The dimensionless parameter $\lambda$ is proportional to the physical confinement length. A density modulation with material wave number $q$ has effective wave number $\lambda q$ on the normalized domain and dimensionless layer spacing $2\pi/(\lambda q)$. At fixed $q$, increasing $\lambda$ places more layers in the same normalized domain, so $u$ becomes more oscillatory and requires higher numerical resolution. The complete rescaling and the dimensionless SmA and SmC energies are given in Appendix~\ref{app:nondim}.

\section{Deep variational framework}
\label{sec:method}

\begin{figure*}[!t]
\centering
\includegraphics[width=\linewidth]{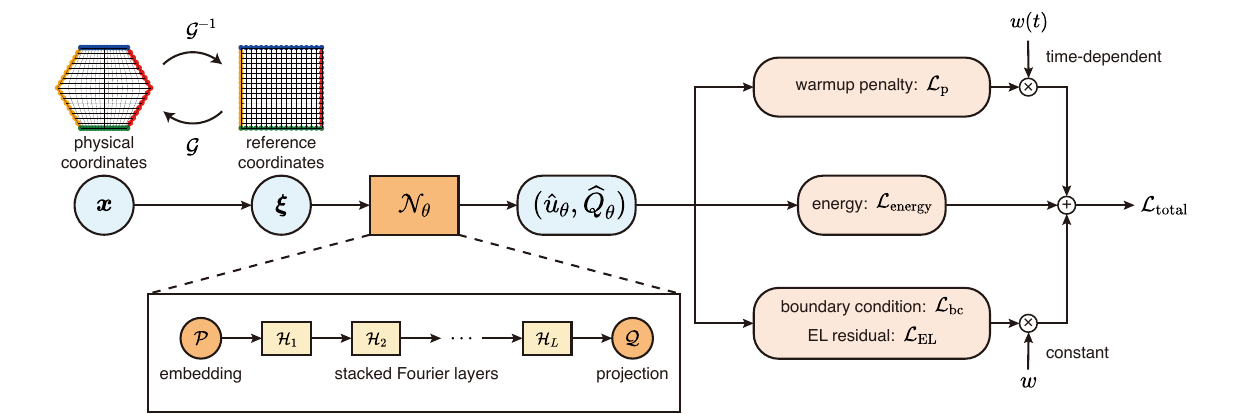}
\caption{Schematic of the deep variational framework (DVF) for computing stable configurations of confined smectic liquid crystals. A physical point $\bm{x}\in\Omega$ is mapped to the reference coordinate $\bm{\xi}\in\widehat{\Omega}$ through $\mathcal{G}^{-1}$, and the neural network $\mathcal{N}_\theta$ is evaluated on a uniform reference grid to produce $(\widehat{u}_\theta,\widehat{\bm{Q}}_\theta)$. Symmetry and tracelessness of $\widehat{\bm{Q}}_\theta$ are imposed by construction. The inset shows the Fourier-based architecture, consisting of an embedding layer, stacked Fourier layers, and a projection layer. The core loss combines the normalized modified Landau--de Gennes (mLdG) free energy $\mathcal{L}_{\mathrm{energy}}$, boundary-condition penalty $\mathcal{L}_{\mathrm{bc}}$, Euler--Lagrange (EL) residual regularization $\mathcal{L}_{\mathrm{EL}}$, and warmup penalty $\mathcal{L}_{\mathrm{p}}$.}
\label{fig:architecture}
\end{figure*}


The mLdG model poses several computational challenges, including complex confinement geometries, higher-order derivatives, and strongly oscillatory smectic layers. To address them, we adapt the DVF to represent the coupled order parameters on a regular reference domain, incorporate physical confinement through coordinate mappings, and use a Fourier-based architecture with a wavelength-guided warmup penalty to facilitate the recovery of layered states. The overall framework is summarized in Fig.~\ref{fig:architecture} and detailed below.

\subsection{Neural representation of the order parameters}
We use a neural network as a trial function for the mLdG order parameters and determine its parameters directly through free-energy minimization. The network is evaluated on a regular reference domain $\widehat{\Omega}$ with coordinate $\bm{\xi}$, and a hat denotes a field represented on this domain. Specifically, the network outputs
\begin{equation}
(\widehat u_\theta,\widehat{\bm Q}_\theta)
=
\mathcal N_\theta(\bm{\xi}),
\label{eq:network_output}
\end{equation}
where $\theta$ denotes the collection of all trainable network parameters. The corresponding fields on the physical domain are defined through the coordinate mapping by
\begin{equation}
u_\theta(\bm{x})
=
\widehat u_\theta(\mathcal G^{-1}(\bm{x})),
\qquad
\bm Q_\theta(\bm{x})
=
\widehat{\bm Q}_\theta(\mathcal G^{-1}(\bm{x})).
\label{eq:physical_fields}
\end{equation}
The coordinate mapping in Eq.~\eqref{eq:physical_fields} is specified in Sec.~\ref{sec:coord_transform}. Substituting $(u_\theta,\bm Q_\theta)$ into the mLdG energy replaces the original field minimization by a finite-dimensional optimization over $\theta$. The loss used to carry out this optimization, including its auxiliary terms, is specified in Sec.~\ref{sec:method_loss}; the optimized network represents the fields of a computed local energy minimum.

To construct these trial fields, we adapt the Fourier-based representation of the DVF developed in Ref.~\cite{xie2026geometry} to the mLdG order parameters. Specifically, the network architecture is
\[
\mathcal N_\theta
=
\mathcal Q\circ \mathcal H_L\circ\cdots\circ \mathcal H_1\circ \mathcal P,
\]
where $\mathcal P$ embeds the input coordinates to a higher-dimensional feature space, $\mathcal H_\ell$ is the $\ell$th Fourier layer, and $\mathcal Q$ projects the final features to $(\widehat u_\theta,\widehat{\bm Q}_\theta)$. Following the Fourier Neural Operator design \cite{li2021fourier}, each Fourier layer combines spectral convolution with a pointwise transformation and nonlinear activation. This architecture provides a Fourier-based representation of the positional and orientational fields. The projection produces $\widehat u_\theta$ and a symmetry-preserving parameterization of $\widehat{\bm Q}_\theta$; the explicit 2D and 3D forms are given in Eq.~\eqref{eq:q_parameterizations}.

\subsection{Coordinate mapping for complex domains}
\label{sec:coord_transform}

To treat different confinement geometries within a common computational setting, we relate the physical domain $\Omega$ to a regular reference domain $\widehat{\Omega}$ through a coordinate transformation. Following the strategy of PhyGeoNet \cite{gao2021phygeonet}, we let $\bm{x}\in\Omega$ and $\bm{\xi}\in\widehat{\Omega}$ denote the physical and reference coordinates, respectively, and introduce a smooth one-to-one map $\bm{x}=\mathcal{G}(\bm{\xi})$ from $\widehat{\Omega}$ to $\Omega$. Equivalently, $\mathcal G^{-1}$ pulls the problem on the complex physical domain back to the regular reference domain. The neural network is then evaluated on a uniform grid in $\widehat{\Omega}$, while the geometry of $\Omega$ is incorporated through $\mathcal G$.

All quantities required by the mLdG energy are evaluated through this mapping. Volume integrals are transformed to $\widehat{\Omega}$ with the corresponding Jacobian factor, and the gradients of $\bm Q$ and the higher-order derivatives of $u$ are first computed on the reference grid and then converted to physical derivatives by the chain rule. The free energy can therefore be evaluated on the regular reference domain and differentiated with respect to the network parameters during optimization. The explicit derivative transformations in 2D and 3D are given in Appendix~\ref{appd:coord}.

\subsection{Free-energy minimization and warmup penalty}
\label{sec:method_loss}

Using the neural representation and coordinate mapping above, we optimize the network parameters within a Deep Ritz formulation \cite{yu2018deep}, in which the mLdG problem is treated as free-energy minimization over the neural trial space. The total loss is
\begin{equation}
\mathcal{L}_{\mathrm{total}}
=
\mathcal{L}_{\mathrm{energy}}
+
w_{\mathrm{bc}} \mathcal{L}_{\mathrm{bc}}
+
w_{\mathrm{EL}} \mathcal{L}_{\mathrm{EL}}
+
w_{\mathrm{p}}(t)\mathcal{L}_{\mathrm{p}}.
\label{eq:loss}
\end{equation}
Here $\mathcal{L}_{\mathrm{energy}}$ is the normalized mLdG free energy and remains the primary variational objective. The term $\mathcal L_{\mathrm{bc}}$ enforces anchoring of $\bm Q$ through a squared Dirichlet mismatch in most cases and a degenerate planar surface potential for 3D tangent anchoring. The explicit forms and their discrete implementation are given in Appendix~\ref{appd:auxloss}.

In a standard variational formulation, the free energy together with the prescribed boundary conditions defines the minimization problem. In our discrete neural implementation, however, minimizing only the energy and boundary penalty can admit grid-scale artifacts localized near defects. Such configurations may have an anomalously low sampled energy while remaining far from a stationary point of the continuum functional. We therefore include the EL residual $\mathcal{L}_{\mathrm{EL}}$, defined in Appendix~\ref{appd:auxloss}, as a small-weight stationarity regularization. It suppresses these under-resolved configurations without replacing the energy as the selection criterion; an unstable critical point can also have a small EL residual.

The principal auxiliary term for recovering the layered structure of the smectic is the warmup penalty $\mathcal{L}_{\mathrm{p}}$. The inherent spectral bias of the network \cite{rahaman2019spectral} makes a smooth, nearly homogeneous $u$ an easily accessible basin during early training, even though the mLdG energy favors modulation at the wave number $\lambda q$. The time-dependent term $w_{\mathrm{p}}(t)\mathcal{L}_{\mathrm{p}}$ promotes this high-frequency content and is defined by
\begin{equation}
\mathcal{L}_{\mathrm{p}}
=
\left(
\sum_{v\in\mathcal{V}}
\bbint_{\mathbb{S}^{d-1}}
\left|
\int_{\Omega}
v(\bm{x})\,\psi_{\bm{\omega}}(\bm{x})
\,\mathrm{d}\bm{x}
\right|
\,\mathrm{d}\bm{\omega}
-
C_{\mathrm p}
\right)^{2}.
\label{eq:warmup_penalty}
\end{equation}
Here $\mathcal{V}
=
\{u\}
\cup
\left\{
\lambda^{-2}\frac{\partial^{2}u}{\partial x_i \partial x_j}
\right\}_{1\le i\le j\le d},$ and $\psi_{\bm{\omega}}(\bm{x})
=
\cos\bigl(q\lambda\, \bm{\omega}\!\cdot\!\bm{x}\bigr)$, where $\bm{\omega}\in\mathbb S^{d-1}$ is a sampled direction and $C_{\mathrm p}>0$ sets the target projection strength. Averaging over $\bm{\omega}$ avoids selecting a preferred layer orientation, and the nonzero target $C_{\mathrm p}$ makes the nonlayered field $u\equiv0$ unfavorable during warmup. Since $w_{\mathrm{p}}(t)\to0$ as $t\to\infty$ (see Appendix~\ref{app:implementation}), the final state is governed solely by the mLdG model. The role of the warmup penalty is tested by the ablation study in Sec.~\ref{sec:square}, and the loss weights and schedules are reported in Appendix~\ref{app:implementation}.


\section{Numerical results}
\label{sec:results}

We now examine mLdG energy minima across a range of 2D and 3D confinements and anchoring conditions. The frustrated mLdG energy landscape supports multiple competing minima under the same physical conditions, as reflected by the distinct configurations obtained from independent initializations. For each parameter setting, we present the minimum with the lowest computed energy, while recognizing that the sampled states do not exhaust the full landscape. To compare the effects of geometry and anchoring within a common ordered regime, we fix $a=-5$, $b=0$, $c=5$, $q=2\pi$, and $B_0=10^{-3}$ for $u$, and take $B=6400$ and $C=3500$ for $\bm Q$. The 2D and 3D computations use the reduced and standard LdG bulk potentials, respectively, with their quadratic coefficients chosen to give the same equilibrium scalar order $s_+=B/C$ \cite{shi2025amodified}. Their explicit forms and parameter choices are given in Appendix~\ref{app:nondim}. Most results are reported at $\lambda^2=30$ and $50$; Appendix~\ref{app:implementation} gives the complete numerical settings.

\subsection{Validation and SmA minima in square confinement}
\label{sec:square}

\begin{figure}[!t]
\centering
\includegraphics[width=\linewidth]{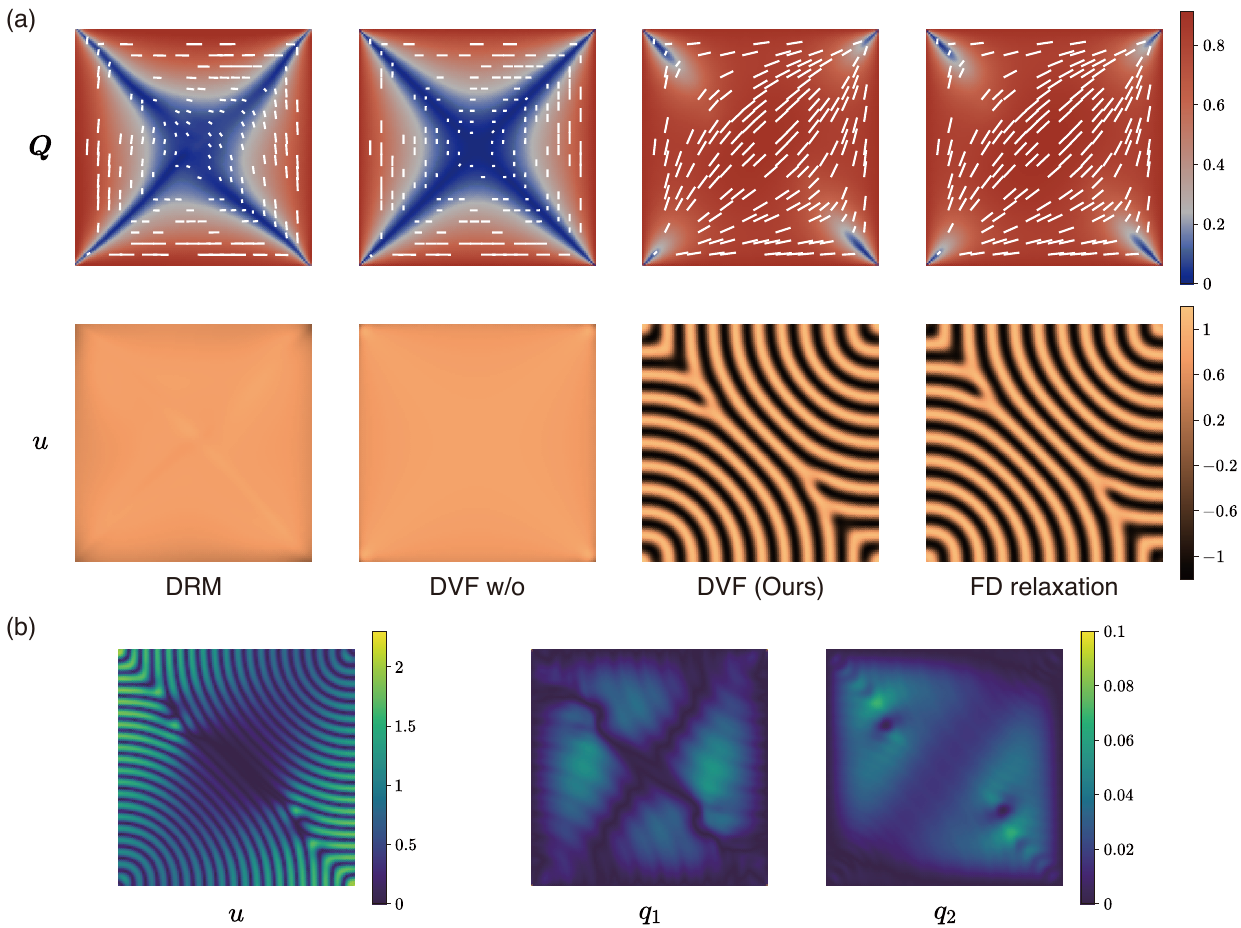}
\caption{Validation and ablation for square-confined smectic-A (SmA) at $\lambda^2=30$ under tangent anchoring. (a) Orientational fields $\bm{Q}$ (top) and positional fields $u$ (bottom) obtained with the Deep Ritz method (DRM), the deep variational framework (DVF) without the warmup penalty (DVF w/o), the full DVF, and finite-difference (FD) relaxation initialized from the full DVF solution. White segments show the director, while the background gives the scalar orientational order $\sqrt{\operatorname{tr}(\bm{Q}^2)/2}$; dark blue indicates reduced order. The DRM and DVF w/o yield nonlayered configurations with well order-reconstruction solution (WORS)-type orientational textures. The full DVF recovers the layered state and its D-like $\bm{Q}$ texture, both of which are retained after FD relaxation. (b) Pointwise differences $|u-u^*|$, $|q_1-q_1^*|$, and $|q_2-q_2^*|$ between the full DVF and FD-relaxed minima, where $q_1$ and $q_2$ are the independent components of the 2D tensor. The rightmost colorbar is shared by $q_1$ and $q_2$.}
\label{fig:baseline}
\makesubfig{fig:baseline_a}
\makesubfig{fig:baseline_b}
\end{figure}

We first verify our framework for SmA in the square confinement $\Omega=[-1,1]^2$ under strong tangent anchoring (i.e., with the director parallel to each edge), which is a representative physical setting \cite{tsakonas2007multistable,cortes2016colloidal,shi2025amodified}. The validation is performed at $\lambda^2=30$. The neural comparison includes the standard DRM as a baseline, the DVF without the warmup penalty (DVF w/o) as a direct ablation, and the full DVF. All three calculations use the same physical parameters and comparable optimization budgets. After the full DVF has converged, its fields are used to initialize a finite-difference (FD) relaxation. The resulting FD-relaxed state is then compared with the DVF solution in both morphology and energy.

The qualitative effect of the warmup penalty is visible in Fig.~\figsubref{fig:baseline}{fig:baseline_a}. Neither the DRM nor DVF w/o develops the high-frequency modulation required for smectic layering. Their orientational fields instead resemble the well order-reconstruction solution (WORS), a characteristic configuration of a square-confined nematic under tangent anchoring in which two diagonal low-order lines divide the square into four sectors with approximately uniform director orientations \cite{canevari2017order,yin2020construction}.

By contrast, the full DVF recovers a well-resolved oscillatory density field and an associated diagonal (D-like) $\bm{Q}$ texture, in which the director is predominantly aligned with one diagonal of the square. Because DVF w/o and the full DVF differ only by the warmup penalty, their distinct outcomes isolate its role in the optimization. The penalty moves the early iterations away from the nearly homogeneous basin and toward the layered smectic branch, while its eventual removal ensures that it does not alter the final minimization objective.

The energy and EL residual in Table~\ref{tab:baseline} distinguish the branches reached by the three neural calculations. The DRM has both a high energy and a large residual, whereas DVF w/o attains a small residual but remains on a higher-energy nonlayered branch. The latter is therefore close to a stationary nonlayered configuration, while the full DVF reaches the substantially lower-energy layered branch. This comparison shows that the warmup penalty primarily changes the basin accessed during optimization rather than simply reducing the residual.

\begin{table}[!t]
\caption{Normalized free energy $\mathcal E$ and EL-residual measure $\mathcal{L}_{\mathrm{EL}}$ for square-confined SmA at $\lambda^2=30$ under tangent anchoring, computed with the three neural settings and with FD relaxation initialized from the full DVF solution.}
\label{tab:baseline}
\begin{ruledtabular}
\begin{tabular}{ccccc}
 & DRM & DVF w/o & DVF (Ours) & FD-relaxed \\
\colrule
$\mathcal E$ & $-1.2767$ & $-1.3483$ & $-3.0336$ & $-3.0358$ \\
EL residual & $1.4\times10^{1}$ & $1.3\times10^{-2}$ & $8.9\times10^{-2}$ & $1.2\times10^{-18}$ \\
\end{tabular}
\end{ruledtabular}
\end{table}

The FD relaxation initialized from the DVF configuration preserves the layered D-like morphology and converges to a stationary state whose Hessian is positive definite. This confirms that the DVF recovers the basin of a stable layered minimum. Using the FD-relaxed state as the reference, Fig.~\figsubref{fig:baseline}{fig:baseline_b} shows localized differences in $\bm Q$ and phase-sensitive differences in the oscillatory field $u$, while the overall D-like layered morphology is retained.

\begin{figure}[!t]
\centering
\includegraphics[width=\linewidth]{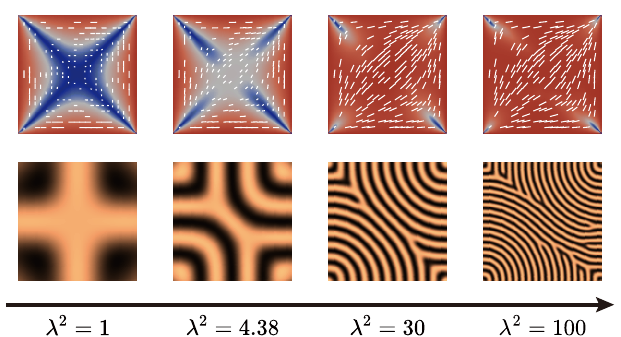}
\caption{SmA minima in a square under tangent anchoring for $\lambda^2=1$, $4.38$, $30$, and $100$. Each column shows $\bm Q$ (top) and $u$ (bottom). The sequence changes from a WORS-like nonlayered minimum, through the onset of layering and reconstruction of the diagonal low-order lines, to D-like layered minima. The visualization and color mapping are the same as in Fig.~\ref{fig:baseline}.}
\label{fig:square_transition}
\end{figure}

We next study the minima in square confinement at $\lambda^2=1$, $4.38$, $30$, and $100$, which resolve the structural transition produced by increasing the effective confinement size. As shown in Fig.~\ref{fig:square_transition}, the $\lambda^2=1$ minimum is WORS-like: two line defects divide the square into four sectors whose directors follow the adjacent edges, while $u$ remains nearly homogeneous. At $\lambda^2=4.38$, the line defects begin to reconstruct, and a layered modulation appears. The minima at $\lambda^2=30$ and $100$ are instead D-like, with a predominantly diagonal bulk director and no extended defect lines across the interior. This transition reflects a change in the relative cost of accommodating the tangent boundary condition. At small $\lambda$, spatial variations carry a high elastic cost relative to bulk ordering, so the boundary alignment influences the entire square and the WORS-like profile provides a low-distortion interpolation between the four edges \cite{canevari2017order}. As $\lambda$ increases, retaining two line defects across the interior becomes unfavorable, and the D-like minima become energetically favorable because they eliminate the interior defects and preserve strong orientational and positional order through most of the bulk \cite{shi2025amodified}.

\subsection{SmA in 2D irregular domains}
\begin{figure}[!t]
\centering
\includegraphics[width=\linewidth]{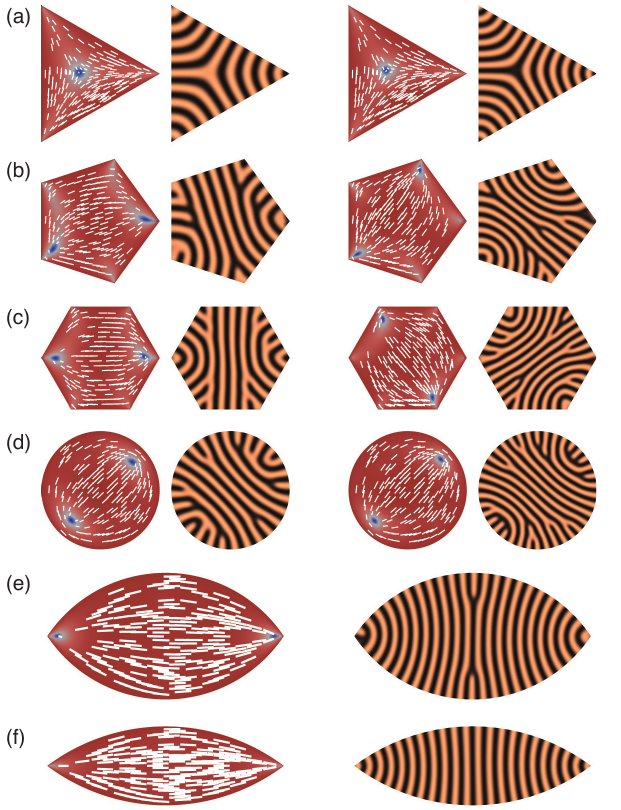}
\caption{SmA minima in 2D irregular domains under tangent anchoring: (a) equilateral triangle, (b) regular pentagon, (c) regular hexagon, (d) disk, (e) wide spindle, and (f) narrow spindle. Rows (a)--(d) show $\lambda^2=30$ (left) and $50$ (right), whereas rows (e) and (f) show only $\lambda^2=30$. Within each group, $\bm Q$ is shown on the left and $u$ on the right. The triangle contains an interior orientational defect and a layer triple junction, whereas the pentagon and hexagon concentrate orientational defects and layer dislocations near selected corners while retaining comparatively regular interior layers. The disk and spindles develop bipolar organizations, with the defects approaching the tips as the spindle narrows. The color mapping is the same as in Fig.~\ref{fig:baseline}.}
\label{fig:polygon}
\makesubfig{fig:polygon_a}
\makesubfig{fig:polygon_b}
\makesubfig{fig:polygon_c}
\makesubfig{fig:polygon_d}
\makesubfig{fig:polygon_e}
\makesubfig{fig:polygon_f}
\end{figure}

We next consider 2D irregular domains under tangent anchoring, as shown in Fig.~\ref{fig:polygon}. The equilateral triangle, regular pentagon, regular hexagon, and disk are studied at $\lambda^2=30$ and $50$, providing a progression from sharply cornered polygons to a smooth boundary. The minima in polygonal confinements are distinguished by how tangent anchoring is accommodated at splay and bend corners. In the equilateral triangle (Fig.~\figsubref{fig:polygon}{fig:polygon_a}), all three vertices are splay-like, and the resulting boundary rotation is balanced by an interior $-1/2$ defect \cite{han2020reduced,rajamanickam2026nematic}. This low-order core anchors a threefold layer junction, while the layers remain nearly regular along the three edges. The minima in pentagonal and hexagonal confinements (Figs.~\figsubref{fig:polygon}{fig:polygon_b} and \figsubref{fig:polygon}{fig:polygon_c}) belong to a different class: two vertices are splay-like, and the remaining vertices are bend-like \cite{han2020reduced,han2021solution}. The orientational mismatch is thereby expelled from the interior and concentrated at the two splay corners. The layer structure responds by terminating or inserting layers near these corners, producing boundary dislocations while preserving a nearly ideal layer arrangement in the central region. These two organizations persist at both values of $\lambda^2$, although symmetry-related choices of the splay corners lead to differently oriented minima.

The disk (Fig.~\figsubref{fig:polygon}{fig:polygon_d}) provides a smooth-boundary counterpart to these polygonal confinements. In the absence of corners that can accommodate the rotation required by tangent anchoring, the orientational field adopts a bipolar configuration with two boundary defects at opposite poles, while the smectic layers bend continuously between them. This confinement-induced localization of orientational and positional defects is consistent with experimental and simulation studies of confined smectics \cite{monderkamp2021topology,wittmann2021particle,wittmann2023colloidal,jull2024curvature}.

To further explore the role of boundary geometry in organizing defects and smectic layers, we turn to the spindle-shaped domains in Figs.~\figsubref{fig:polygon}{fig:polygon_e} and \figsubref{fig:polygon}{fig:polygon_f}, which connect the preceding confinement geometries to tactoid-like shapes. Such pointed domains commonly arise during isotropic--nematic coexistence, where elasticity, anchoring, and anisotropic interfacial energy jointly determine the droplet shape and director field \cite{prinsen2003shape,kim2013morphogenesis}. The wide spindle in Fig.~\figsubref{fig:polygon}{fig:polygon_e} exhibits a bipolar-like defect pair accompanied by pronounced curvature and distortion of the interior layers. As the spindle becomes narrower (Fig.~\figsubref{fig:polygon}{fig:polygon_f}), regions of strong director variation become increasingly localized near the high-curvature tips, allowing the elongated central region to sustain a nearly uniform director and straighter, more regularly spaced layers. This comparison suggests that pointed confinement can concentrate orientational frustration near the tips while promoting a more uniformly ordered smectic interior.

\subsection{SmA in 3D confinements}

\begin{figure}[!t]
\centering
\includegraphics[width=\linewidth]{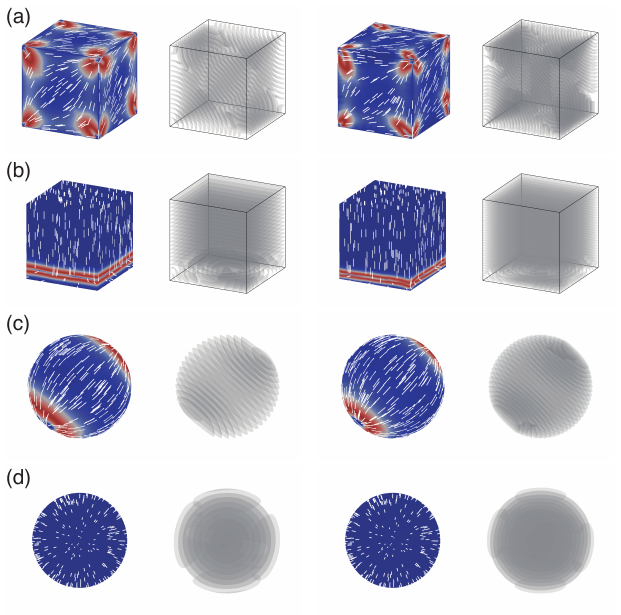}
\caption{SmA minima in 3D under different geometries and anchoring conditions: (a) cube with tangent anchoring, (b) cube with TFCD-inducing anchoring, (c) sphere with tangent anchoring, and (d) sphere with homeotropic anchoring. In each row, the left and right groups correspond to $\lambda^2=30$ and $50$, respectively, and show $\bm Q$ together with $u$. White director lines are superimposed on the biaxiality $\beta=1-6(\operatorname{tr}\bm Q^3)^2/(\operatorname{tr}\bm Q^2)^3$; regions of enhanced biaxiality, which highlight defect cores and order-reconstruction regions, appear red. Smectic layers are represented by zero isosurfaces of $u$. The four cases exhibit corner- and edge-localized distortions, a near-substrate order-reconstruction layer, bipolar curved layers, and concentric shells, respectively.}
\label{fig:sma3d}
\makesubfig{fig:sma_a}
\makesubfig{fig:sma_b}
\makesubfig{fig:sma_c}
\makesubfig{fig:sma_d}
\end{figure}

We next extend the analysis of confined SmA minima from two to three dimensions and examine how the equilibrium structures respond to changes in confinement geometry and surface anchoring. We begin with tangent anchoring in the cube and sphere (Figs.~\figsubref{fig:sma3d}{fig:sma_a} and \figsubref{fig:sma3d}{fig:sma_c}). In the cube, the director is approximately aligned with a body diagonal throughout much of the interior, whereas stronger distortions develop near the boundary. A nearly uniform bulk orientation cannot satisfy tangent anchoring simultaneously on all six faces, and the resulting mismatch is therefore accommodated primarily near the edges and corners. Correspondingly, the smectic layers remain comparatively regular in the interior and become distorted mainly near the faceted boundary. This localization of orientational distortion is consistent with the multistable tangent-anchored textures reported for nematic cuboids \cite{shi2024multistability}. The spherical domain exhibits a different organization. Under tangent anchoring, the $\bm Q$ field adopts a bipolar texture with two antipodal low-order regions, while the smectic layers form a smoothly curved stack connecting the two poles. The strongest distortions are concentrated near these polar regions. In contrast to the cube, the smooth spherical boundary provides no edges or corners at which the anchoring mismatch can be localized, and the frustration is instead organized around the bipolar defect pair. Similar bipolar structures have been observed in confined nematic droplets and in particle simulations of spherical smectics \cite{lopez2009topological,whitmer2013nematic,zhou2023defect}.

Changing the spherical boundary condition from tangent to homeotropic anchoring leads to a qualitatively different minimum (Fig.~\figsubref{fig:sma3d}{fig:sma_d}). The director is now approximately radial throughout the domain. Because the SmA layer normal preferentially follows the director, the smectic layers close into concentric, onion-like shells rather than forming the curved bipolar stack found under tangent anchoring. Similar radial orientational order and spherical SmA layering have been reported in particle simulations \cite{emerson1995monte,puschel2017impact}. The comparison between Figs.~\figsubref{fig:sma3d}{fig:sma_c} and \figsubref{fig:sma3d}{fig:sma_d} therefore shows that the same spherical confinement can support markedly different layered organizations: tangent anchoring produces a bipolar texture with distortions concentrated near two surface regions, whereas homeotropic anchoring favors a radially organized state with closed concentric layers.

We finally consider the competing anchoring conditions associated with toroidal focal conic domains (TFCDs). In a canonical TFCD, the smectic layers form toroidal structures organized around axial and circular focal lines, mediating between vertical alignment in the bulk and radial in-plane alignment at the substrate \cite{xia2021structural,gim2017morphogenesis,kim2012thermally}. To probe whether a related structure emerges in the present model, we impose vertical alignment at the top surface of the cube and radial alignment at the bottom (Fig.~\figsubref{fig:sma3d}{fig:sma_b}). In both computed minima, the upper portion of the domain retains an almost vertical director together with a nearly uniform stack of horizontal layers. The reorientation required by the bottom boundary remains confined to a narrow region adjacent to the substrate rather than extending through the full height of the cube. Within this near-surface region, the magnitude of $\bm Q$ decreases and its biaxiality increases around a localized low-order core, indicating a biaxial order-reconstruction zone \cite{ambrozic2007defect}. The smectic layers simultaneously develop focal-conic-like curvature around the core, thereby providing a local transition between the radial bottom alignment and the horizontal bulk stack while leaving the upper layers nearly undistorted. This near-substrate reconstruction is also observed in computations at smaller system sizes and persists after FD relaxation.

\subsection{SmC in spherical confinement}
SmC differs from SmA by a finite preferred angle between the director and the layer normal. In conventional confined SmC cells, the onset of this tilt is often accompanied by a reduction in the equilibrium layer spacing across the SmA--SmC transition. When such contraction is frustrated by surface constraints, the layers can reorganize into oppositely tilted segments, giving rise to the characteristic chevron structures observed experimentally and in molecular simulations \cite{rieker1987chevron,webster2003molecular}. Here we separate this conventional layer-contraction mechanism from the effect of the director--layer tilt itself. We therefore retain the same spherical domain, tangent anchoring, material parameters, and preferred wave number $q$ as in the corresponding SmA calculation, but impose a preferred angle $\theta_0=\pi/6$ through the SmC coupling \cite{xia2024simple}.

\begin{figure}[!t]
\centering
\includegraphics[width=\linewidth]{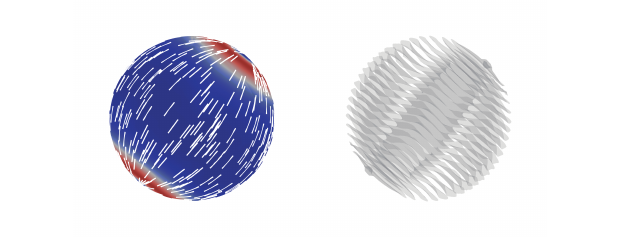}
\caption{SmC minimum in a spherical domain under tangent anchoring at $\theta_0=\pi/6$. The $\bm Q$ field is shown by white director lines superimposed on a biaxiality map, and the layers are represented by zero isosurfaces of $u$. The bipolar orientational texture is accompanied by surface-localized, chevron-like bending of the layers.}
\label{fig:smc}
\end{figure}

The resulting minimum is shown in Fig.~\ref{fig:smc}. Compared with the tangent-anchored SmA state in Fig.~\figsubref{fig:sma3d}{fig:sma_c}, the large-scale orientational organization changes only weakly: both states exhibit a bipolar $\bm Q$ texture with two antipodal low-order regions. Their layer morphologies, however, are markedly different. The SmA state consists of a smoothly curved, approximately parallel stack, whereas the SmC layers undergo pronounced bending and reorientation near the spherical boundary, producing a surface-localized chevron-like pattern. Because the two calculations have the same preferred wave number $q$, this restructuring cannot be attributed to an imposed change in the equilibrium layer spacing and instead originates from the preferred SmC director--layer tilt.

This distinction can be understood geometrically. Tangent anchoring constrains the director to the local tangent plane of the sphere, while the SmC coupling favors a finite angle $\theta_0$ between the director and the layer normal. Since the tangent plane varies continuously over the curved boundary, a single SmA-like layer stack cannot accommodate this preferred angle everywhere without introducing additional distortion. The incompatibility is therefore relieved primarily through spatially varying bending and reorientation of the layers near the surface, while the bipolar orientational texture is largely retained. The resulting structure is thus chevron-like in morphology, but differs from the conventional planar chevron mechanism driven by layer contraction. This curvature-induced response is qualitatively consistent with studies of tangent-anchored smectic shells, where SmC tilt similarly provides an additional route for accommodating geometric frustration through layer bending \cite{manyuhina2015thick,sharma2024smectic}.

\section{Discussion and conclusion}
\label{sec:conclusion}
This work develops a unified computational framework for stable configurations of confined smectics based on an adapted DVF that combines a Fourier representation, coordinate mappings, and a warmup penalty for wavelength guidance. The square-domain ablation shows that the warmup stage is essential for reliably accessing layered mLdG states in the tested regime, while subsequent FD relaxation confirms that the recovered configurations persist as minima under an independent spatial discretization.

Across the computed states, confinement geometry and anchoring organize the large-scale $\bm Q$ field, while director--layer coupling links this orientational organization to the resulting layer morphology. The calculations recover several established structures, including WORS-like and D-like square states, bipolar spherical textures, and TFCD-like and chevron-like morphologies \cite{canevari2017order,shi2025amodified,xia2021structural,rieker1987chevron}. These results demonstrate how geometry, anchoring, and director--layer coupling jointly control defect localization and layer morphology in confined smectics.

The present study is limited to static states in fixed domains and to a finite range of parameters and boundary conditions. The DVF could be adapted to other free-energy models with spatially modulated order, including those for cholesteric liquid crystals \cite{fialho2017effect} and other layered materials. Its geometric scope could likewise be extended from the solid confinements considered here to shell geometries with inner and outer boundaries, particularly spherical shells \cite{manyuhina2015thick,sharma2024smectic}. Beyond prescribed boundaries, diffuse-interface LdG models already couple nematic order to a deformable interface and recover equilibrium droplet morphologies \cite{wu2025diffuse,shi2026analyzing}. A future neural variational formulation could represent both the order parameters and the free boundary by trainable functions and determine them simultaneously through minimization of a common free-energy objective.

\section*{Acknowledgments}
L.Z. was supported by the National Natural Science Foundation of China (No. 12225102, T2321001, and 12288101) and the National Key Research and Development Program of China 2024YFA0919500. Y.H. was supported by Young Faculty Development Fund of the School of Mathematics at Renmin University of China and the Renmin University of China New Faculty's fund. 

\section*{Data availability}
The codes that support the findings of this article are available from the authors upon reasonable request.

\appendix
\section{Nematic bulk potentials and nondimensionalization}
\label{app:nondim}

We first record the tensor parameterizations and dimension-dependent nematic bulk potentials used in the computations. Symmetry and tracelessness are enforced by writing
\begin{equation}
\begin{aligned}
\bm Q^{\mathrm{2D}}
&=
\begin{pmatrix}
q_1&q_2\\
q_2&-q_1
\end{pmatrix},\\[3pt]
\bm Q^{\mathrm{3D}}
&=
\begin{pmatrix}
q_1-q_3&q_2&q_4\\
q_2&-q_1-q_3&q_5\\
q_4&q_5&2q_3
\end{pmatrix}.
\end{aligned}
\label{eq:q_parameterizations}
\end{equation}
Thus, the network uses two scalar fields for $\bm Q$ in 2D and five in 3D. In 3D, the standard LdG potential is
\begin{equation*}
\begin{aligned}
f_{\mathrm{bn}}^{\mathrm{3D}}(\bm Q)
&=
\frac{A}{2}\operatorname{tr}(\bm Q^2)
-\frac{B}{3}\operatorname{tr}(\bm Q^3)
+\frac{C}{4}\bigl(\operatorname{tr}(\bm Q^2)\bigr)^2,\\
s_+^{\mathrm{3D}}
&=
\frac{B+\sqrt{B^2-24AC}}{4C}.
\end{aligned}
\end{equation*}
Here $A=\alpha_1(T-T_1^\ast)$ is temperature dependent, while $B,C>0$ are material parameters. In 2D, $\operatorname{tr}(\bm Q^3)=0$, and choosing $A=-B^2/(2C)$ reduces the bulk potential to
\begin{equation*}
f_{\mathrm{bn}}^{\mathrm{2D}}(\bm Q)=
-\frac{B^2}{4C}\operatorname{tr}(\bm Q^2)
+\frac{C}{4}\bigl(\operatorname{tr}(\bm Q^2)\bigr)^2,
\quad
s_+^{\mathrm{2D}}=\frac{B}{C}.
\end{equation*}
The reduced potential is used in all 2D computations. Its difference from the 3D bulk potential results from reducing the tensor itself to the two-component form above, not merely from restricting the spatial domain to two dimensions. Since the chosen value of $A$ has already been substituted, it does not appear as an independent coefficient in this expression. The resulting quartic potential is the standard reduced Landau--de Gennes (rLdG) form used for 2D nematic equilibria \cite{han2020reduced}.

We next summarize the rescaling used to obtain the dimensionless energies in the main text. Let $L$ be the characteristic physical length of $\Omega$, and set $\bar{\bm{x}}=\bm{x}/L$, so that $\bar{\Omega}=L^{-1}\Omega$, $\nabla=L^{-1}\bar{\nabla}$, $\Delta=L^{-2}\bar{\Delta}$, $D^2=L^{-2}\bar{D}^2$, and $\mathrm{d}\bm{x}=L^d\,\mathrm{d}\bar{\bm{x}}$. Measuring the free energy in units of $KL^{d-2}$ gives $\bar{E}=E/(KL^{d-2})$. We further introduce $\bar{\lambda}^2=2CL^2/K$, $\bar{a}=a/(2C)$, $\bar{c}=c/(2C)$, $\bar{q}^2=Kq^2/(2C)$, and $\bar{B}_0=2B_0C/K^2$, and set $b=0$ as in the main text.

Substituting these relations into the dimensional free-energy functional yields the following dimensionless energies for the SmA and SmC phases:
\begin{equation}
\begin{aligned}
\bar E_{\mathrm{SmA}}(\bm Q,u)
={}&
\int_{\bar\Omega}\Bigg\{
\bar\lambda^2
\bigl[f_u(u)+f_Q(\bm Q)\bigr]
+\frac{1}{2}|\bar\nabla\bm Q|^2
\\
&\quad
+\frac{\bar B_0}{\bar\lambda^2}
\left|
\bar D^2u
+\bar\lambda^2\bar q^2
\left(
\frac{\bm Q}{s_+}
+\frac{\bm I_d}{d}
\right)u
\right|^2
\Bigg\}
\,\mathrm{d}\bar{\bm x},
\end{aligned}
\label{eq:app_energy_sma}
\end{equation}

\begin{equation}
\begin{aligned}
&\bar E_{\mathrm{SmC}}(\bm Q,u)
=
\int_{\bar\Omega}\Bigg\{
\bar\lambda^2
\bigl[f_u(u)+f_Q(\bm Q)\bigr]
+\frac{1}{2}|\bar\nabla\bm Q|^2
\\
&\qquad \qquad \qquad \qquad
+\frac{\bar B_0}{\bar\lambda^2}
\left(
\bar\Delta u
+\bar\lambda^2\bar q^2u
\right)^2
\\
&
+\frac{\bar B_0}{\bar\lambda^2}
\Bigg[
\operatorname{tr}\left(
\bar D^2u
\left(
\frac{\bm Q}{s_+}
+\frac{\bm I_d}{d}
\right)
\right)+\bar\lambda^2\bar q^2u
\cos^2\theta_0
\Bigg]^2
\Bigg\}
\,\mathrm{d}\bar{\bm x},
\end{aligned}
\label{eq:app_energy_smc}
\end{equation}
where
\begin{equation*}
\begin{aligned}
&f_u(u)
=
\frac{\bar a}{2}u^2
+\frac{\bar c}{4}u^4,\\
f_Q(\bm Q)
=
\frac{A}{4C}&\operatorname{tr}(\bm Q^2)
-\frac{B}{6C}\operatorname{tr}(\bm Q^3)
+\frac{\bigl[\operatorname{tr}(\bm Q^2)\bigr]^2}{8}.
\end{aligned}
\end{equation*}
The displayed nematic bulk term is used in 3D. For the 2D rLdG computations, it is replaced by
\[
-\frac{B^2}{8C^2}\operatorname{tr}(\bm Q^2)
+\frac{\bigl(\operatorname{tr}(\bm Q^2)\bigr)^2}{8},
\]
which is the dimensionless form of $f_{\mathrm{bn}}^{\mathrm{2D}}$. The implementation minimizes the further normalized energy
\begin{equation*}
\mathcal E=\frac{\bar E}{\bar\lambda^2}.
\end{equation*}
For each fixed $\bar\lambda$, this positive rescaling leaves the energy minima unchanged and keeps the reported values on a comparable numerical scale over the range of $\bar\lambda$ considered here. Bars are dropped in the main text, and $\mathcal L_{\mathrm{energy}}=\mathcal E$ is used in all loss functions and reported energy values.

\section{Coordinate mapping and transformed derivatives}
\label{appd:coord}
To treat irregular confinement geometries within a unified computational setting, we map the regular reference domain $\widehat{\Omega}=[0,1]^d$, with $d=2,3$, to the physical domain $\Omega$ through an invertible transformation
\[
\bm{x}=\mathcal{G}(\bm{\xi}),
\qquad
\bm{x}\in\Omega,\quad \bm{\xi}\in\widehat{\Omega}.
\]
Mapped geometries are evaluated on this uniform reference grid, while the square and cube are evaluated directly on uniform grids in $[-1,1]^d$. The two settings differ only by an affine change of reference coordinates. Physical derivatives for mapped geometries are recovered through the Jacobian of the transformation, which is assumed to remain nonsingular at the quadrature points.

Following the coordinate-transformation idea used in Ref.~\cite{gao2021phygeonet}, the mapping is defined by prescribing a one-to-one correspondence between $\partial\Omega$ and $\partial\widehat{\Omega}$. The inverse coordinates $\bm{\xi}(\bm{x})=\mathcal{G}^{-1}(\bm{x})$ may be viewed as harmonic coordinates satisfying
\begin{equation*}
\Delta_{\bm{x}} \xi_k = 0,
\qquad k=1,\dots,d,
\end{equation*}
subject to the prescribed boundary correspondence. In practice, it is more convenient to solve for the forward map $\bm{x}=\mathcal{G}(\bm{\xi})$. Writing $\bm{x}=(x_1,\dots,x_d)^T$, the corresponding equations can be written as
\begin{equation*}
\operatorname{tr}\!\left(J^{-T} D_{\bm{\xi}}^2 x_i J^{-1}\right)=0,
\qquad i=1,\dots,d,
\end{equation*}
where $J(\bm{\xi})=\partial \bm{x}/\partial \bm{\xi}$ is the Jacobian matrix and $D_{\bm{\xi}}^2 x_i$ is the Hessian of $x_i$ with respect to the reference coordinates.

Once the map $\bm{x}=\mathcal{G}(\bm{\xi})$ is known, all geometric quantities are evaluated on the reference grid. The Jacobian $J$ is obtained from the mapped coordinates, and its inverse $(J^{-1})_{ki}=\partial \xi_k/\partial x_i$ gives the first-order geometric coefficients. For a scalar field $f(\bm{x})$, let
\[
\widehat{f}(\bm{\xi}) = f(\mathcal{G}(\bm{\xi}))
\]
denote its pullback to the reference domain. The same transformation is applied componentwise to the tensor field $\bm{Q}$, and the volume element satisfies $\mathrm{d}\bm{x}=|\det J|\,\mathrm{d}\bm{\xi}$.

The first-order derivatives in the physical domain are then given by
\begin{equation}
\nabla_{\bm{x}} f = J^{-T}\nabla_{\bm{\xi}} \widehat{f}.
\label{eq:first_derivative_general}
\end{equation}
Thus, derivatives are evaluated on the uniform reference grid and converted to the physical domain through the inverse Jacobian.

For second-order derivatives, the nonlinearity of the coordinate map also enters. Applying the chain rule twice gives
\begin{equation}
D_{\bm{x}}^2 f
=
J^{-T}(D_{\bm{\xi}}^2 \widehat{f})J^{-1}
+
\sum_{k=1}^d
\frac{\partial \widehat{f}}{\partial \xi_k}\,
D_{\bm{x}}^2 \xi_k.
\label{eq:hessian_transform_general}
\end{equation}
The first term is the standard transformation of the Hessian, while the second term accounts for the local curvature of the mapping.

The second-order geometric coefficients are obtained from the identity
\[
\xi_k(\mathcal{G}(\bm{\xi}))=\xi_k,
\qquad k=1,\dots,d.
\]
Differentiating twice with respect to $\bm{\xi}$ yields
\begin{equation}
D_{\bm{x}}^2 \xi_k
=
-\,J^{-T}
\left(
\sum_{l=1}^d (J^{-1})_{kl}\, D_{\bm{\xi}}^2 x_l
\right)
J^{-1},
\;\; k=1,\dots,d.
\label{eq:hessian_inverse_general}
\end{equation}
Therefore, once the forward map is available, Eqs.~\eqref{eq:first_derivative_general}--\eqref{eq:hessian_inverse_general} provide all first- and second-order derivatives in the physical domain in terms of quantities defined on the reference grid.

In the present work, these formulas are used to evaluate the gradient terms of $\bm{Q}$ and the higher-order derivatives of $u$ within the same variational framework for both regular and irregular confinement geometries.

\section{Boundary conditions and auxiliary loss terms}
\label{appd:auxloss}

\subsection{Boundary conditions}

Let $\Gamma_{\mathrm{bc}}\subseteq\partial\Omega$ denote the anchored part of the boundary. Most calculations prescribe $\bm Q_{\mathrm{bc}}$ through
\begin{equation}
\mathcal L_{\mathrm{bc}}
=
\int_{\Gamma_{\mathrm{bc}}}
|\bm Q-\bm Q_{\mathrm{bc}}|^2\,\mathrm dS.
\label{eq:dirichlet_q_bc}
\end{equation}
For 2D tangent anchoring and 3D homeotropic anchoring on the sphere, the targets are
\begin{equation*}
\begin{aligned}
\bm Q_{\mathrm{bc}}^{\mathrm{2D}}
&=s_+\left(\bm t\otimes\bm t-\frac{\bm I_2}{2}\right),\\
\bm Q_{\mathrm{bc}}^{\mathrm{hom}}
&=s_+\left(\bm\nu\otimes\bm\nu-\frac{\bm I_3}{3}\right).
\end{aligned}
\end{equation*}
where $\bm t$ is the local unit tangent, applied edgewise on polygons \cite{han2020reduced,shi2025amodified}, and $\bm\nu$ is the outward radial normal on the sphere. The TFCD-inducing cube uses the same penalty only on its top and bottom faces, with
\begin{equation*}
\begin{aligned}
\bm Q_{\mathrm{top}}
&=s_+\left(\bm e_z\otimes\bm e_z-\frac{\bm I_3}{3}\right),\\
\bm Q_{\mathrm{bot}}
&=s_+\left(\bm e_r\otimes\bm e_r-\frac{\bm I_3}{3}\right),
\end{aligned}
\end{equation*}
where $\bm e_z=(0,0,1)$ is the vertical unit vector and $\bm e_r=(\cos\varphi,\sin\varphi,0)$ is the in-plane radial direction on the bottom face. Here $\Gamma_{\mathrm{bc}}$ consists of the top and bottom faces; the four lateral faces are left unanchored \cite{gim2017morphogenesis,xia2021structural}.

The 3D tangent cases instead use the degenerate planar surface loss
\begin{equation}
\mathcal L_{\mathrm{bc}}
=
\int_{\partial\Omega}
\left|
\left(\bm Q+\frac{s_+}{3}\bm I_3\right)\bm\nu(\bm x)
\right|^2\,\mathrm dS.
\label{eq:tangent_bc}
\end{equation}
This places a uniaxial director in the tangent plane without selecting an in-plane direction \cite{shi2024multistability}. It is used for the tangent-anchored cube and the SmA and SmC spheres.

In contrast, no boundary term is imposed on $u$. Since the mLdG energy depends on $D^2u$, its natural boundary condition is
\begin{equation*}
\left.
\frac{\mathrm d}{\mathrm d\varepsilon}
E\bigl(\bm Q,u+\varepsilon v\bigr)
\right|_{\varepsilon=0}
=0
\qquad
\text{for every }v\in H^2(\Omega),
\end{equation*}
where the traces of $v$ and $\partial_{\bm\nu}v$ are unrestricted. Thus neither $u$ nor $\partial_{\bm\nu}u$ is prescribed at $\partial\Omega$ \cite{xia2023variational,shi2025amodified}; $\bm Q$ is anchored explicitly, while the boundary behavior of $u$ is determined variationally.

\subsection{Euler--Lagrange residual penalty}

For completeness, we summarize the EL residual penalty introduced in Sec.~\ref{sec:method_loss}. The formulas below correspond to the dimensionless energies $\bar E$ in Eqs.~\eqref{eq:app_energy_sma} and \eqref{eq:app_energy_smc}, using the bar-dropped parameter notation of the main text. It is convenient to define the equivalent stationary operators
\begin{equation*}
\mathcal R_{\bm Q}
=-\frac{\delta\bar E}{\delta\bm Q},
\qquad
\mathcal R_u
=\frac{\lambda^2}{2B_0}\frac{\delta\bar E}{\delta u}.
\end{equation*}
Their common zero set is the EL system. Writing the parameterizations in Eq.~\eqref{eq:q_parameterizations} as $\bm Q=\bm Q(q_1,\ldots,q_m)$, with $m=2$ in 2D and $m=5$ in 3D, and using $\mathcal E=\bar E/\lambda^2$, the implemented residuals are
\begin{equation}
r_\alpha
=-\lambda^{-2}\mathcal R_{\bm Q}:\frac{\partial\bm Q}{\partial q_\alpha},
\qquad
r_u
=2B_0\lambda^{-4}\mathcal R_u.
\label{eq:implemented_residuals}
\end{equation}
The regularization is their squared Euclidean norm,
\begin{equation*}
\mathcal L_{\mathrm{EL}}
=
\int_\Omega
\left(
\sum_{\alpha=1}^{m}|r_\alpha|^2+|r_u|^2
\right)\,\mathrm{d}\bm x.
\end{equation*}
Thus, $\mathcal L_{\mathrm{EL}}$ measures stationarity with respect to the scalar fields produced by the network. For mapped geometries, the derivatives and integral are evaluated using Appendix~\ref{appd:coord}.

\paragraph*{SmA phase.}
For the $d$-dimensional SmA free energy, the orientational EL residual is
\begin{align*}
\mathcal{R}_{\bm{Q}}^{\text{SmA}}
={}& \Delta \bm{Q}
- 2\lambda^2 B_0 q^4 \frac{\bm{Q}}{s_+^2} u^2 \\
&- \frac{2B_0 q^2}{s_+}
\left( u D^2u - \frac{\operatorname{tr}(u D^2u)}{d}\bm{I}_d \right) \\
&\!\!\!\!\!\!\!\!\!\!\!\! - \lambda^2
\left[
\frac{A}{2C}\bm{Q}
- \frac{B}{2C}
\left( \bm{Q}^2 - \frac{\operatorname{tr}(\bm{Q}^2)}{d}\bm{I}_d \right)
+ \frac{\operatorname{tr}(\bm{Q}^2)}{2}\bm{Q}
\right],
\end{align*}
where the final bulk contribution is replaced in 2D by
\[
-\lambda^2
\left(
-\frac{B^2}{4C^2}
+\frac{\operatorname{tr}(\bm Q^2)}{2}
\right)\bm Q,
\]
in accordance with the rLdG potential.
The positional EL residual is
\begin{align*}
\mathcal{R}_u^{\text{SmA}}
={}& \Delta^2 u
+ \lambda^4 \left( \frac{a}{2B_0}u + \frac{c}{2B_0}u^3 \right) \\
&+ \lambda^2 \nabla \cdot
\left[
\nabla \cdot
\left(
q^2 \left(\frac{\bm{Q}}{s_+} + \frac{\bm{I}_d}{d}\right) u
\right)
\right] \\
&+ \lambda^4
\left| q^2 \left(\frac{\bm{Q}}{s_+} + \frac{\bm{I}_d}{d}\right) \right|^2 u \\
&+ \lambda^2 D^2 u :
\left[
q^2 \left(\frac{\bm{Q}}{s_+} + \frac{\bm{I}_d}{d}\right)
\right].
\end{align*}

\paragraph*{SmC phase.}
For the SmC free energy, it is convenient to introduce the auxiliary scalar field
\begin{equation*}
\mathcal{W}
=
\operatorname{tr}
\left[
D^2u
\left(
\frac{\bm{Q}}{s_+}+\frac{\bm{I}_d}{d}
\right)
\right]
+
\lambda^2 q^2 u \cos^2\theta_0.
\end{equation*}
The corresponding residuals are then written as
\begin{align*}
\mathcal{R}_{\bm{Q}}^{\text{SmC}}
={}& \Delta \bm{Q}
- \frac{2B_0}{\lambda^2 s_+}\mathcal{W}
\left( D^2 u - \frac{\Delta u}{d}\bm{I}_d \right) \\
&\!\!\!\!\!\!\!\!\!\!\!\! - \lambda^2
\left[
\frac{A}{2C}\bm{Q}
- \frac{B}{2C}
\left( \bm{Q}^2 - \frac{\operatorname{tr}(\bm{Q}^2)}{d}\bm{I}_d \right)
+ \frac{\operatorname{tr}(\bm{Q}^2)}{2}\bm{Q}
\right],
\end{align*}
and
\begin{align*}
\mathcal{R}_u^{\text{SmC}}
={}& \Delta^2 u
+ 2\lambda^2 q^2 \Delta u
+ \lambda^4 q^4 u \\
&+ \lambda^4 \left( \frac{a}{2B_0}u + \frac{c}{2B_0}u^3 \right) \\
&\!\!\!\! + \nabla \cdot
\left[
\nabla \cdot
\left(
\mathcal{W}
\left( \frac{\bm{Q}}{s_+} + \frac{\bm{I}_d}{d} \right)
\right)
\right]
+ \lambda^2 q^2 \mathcal{W} \cos^2\theta_0.
\end{align*}

Equation~\eqref{eq:implemented_residuals} converts these compact stationary operators to the componentwise residuals used in the code and in Table~\ref{tab:baseline}.

\section{Implementation details}
\label{app:implementation}

Unless otherwise stated, all DVF computations use four Fourier layers with 64 hidden channels, Gaussian error linear unit (GELU) activation, and 16 retained Fourier modes in each direction \cite{hendrycks2016gaussian}. We use 129 grid points per direction in 2D and 65 in 3D, with all fields and derivatives evaluated in double precision. AdamW is applied with a learning rate of $10^{-3}$ and a weight decay of $10^{-2}$ for $5000$ iterations in 2D and $10000$ in 3D \cite{loshchilov2019decoupled}; the gradient norm is clipped at $1$. Each 2D setting is computed from 100 independent initializations, while each 3D setting uses 50 independent initializations; the figures show the lowest-energy minimum found for each setting.

The loss weights are fixed as $w_{\mathrm{bc}}=100$ and $w_{\mathrm{EL}}=0.1$ in all experiments. The warmup penalty is applied with the time-dependent weight
\begin{equation}
w_{\mathrm{p}}(t)=
\begin{cases}
w_{\mathrm{p}}^{\max}, & t<250,\\
w_{\mathrm{p}}^{\max}(500-t)/250, & 250\le t<500,\\
0, & t\ge 500,
\end{cases}
\label{eq:warmup_schedule}
\end{equation}
with $(C_{\mathrm p},w_{\mathrm{p}}^{\max})=(50,0.01)$ in 2D and $(80,0.02)$ in 3D. The directional average in $\mathcal L_{\mathrm p}$ is estimated using 32 independently sampled directions per iteration.

For the 3D computations, we additionally employ temporary orientational guidance during the first 500 iterations, analogous to that used for phases with narrow attraction basins in Ref.~\cite{xie2026geometry}. The objective in Eq.~\eqref{eq:loss} is augmented by $w_{\mathrm d}(t)\mathcal L_{\mathrm d}$, with
\begin{equation*}
\mathcal L_{\mathrm d}
=
\frac{1}{|\Omega|}
\int_{\Omega}
\left|{\bm Q}_\theta-\bm Q_{\mathrm{ref}}\right|^2
\,\mathrm{d}\bm{x}.
\end{equation*}
The reference field is constructed as $\bm Q_{\mathrm{ref}}=s_+(\bm n_{\mathrm{ref}}\otimes\bm n_{\mathrm{ref}}-\bm I_3/3)$. The tangent-anchored cube and the tangent-anchored SmA and SmC spheres use the uniform direction $\bm n_{\mathrm{ref}}=(1,1,1)/\sqrt{3}$, whereas the homeotropic sphere uses the radial direction $\bm n_{\mathrm{ref}}=\bm x/|\bm x|$. The TFCD calculation uses the regularized toroidal director ansatz adopted in Ref.~\cite{xia2021structural}. No reference field is imposed on $u$. The weight $w_{\mathrm d}(t)$ has a maximum value of $15$ and decreases to zero after iteration $500$.

The DRM baseline uses four residual multilayer-perceptron blocks with 64 hidden channels and the same optimizer, precision, and 5000-iteration budget. At each iteration, it samples 2000 interior points and 200 points on each edge. For the FD cross-check in Sec.~\ref{sec:square}, the DVF fields are interpolated to the square grid with spacing $h=2/128$ and relaxed by Barzilai--Borwein gradient descent \cite{barzilai1988two} until the squared discrete gradient norm is below $10^{-8}$. Local stability is assessed by testing the positive definiteness of the Hessian at the converged state.

\bibliography{references}

\end{document}